\documentclass[aps, amsmath, amssymb, 10pt, prl, twocolumn, longbibliography, superscriptaddress]{revtex4-1}

\usepackage{algcompatible}
\usepackage{amsmath}
\usepackage{amsthm}
\usepackage{bm}
\usepackage{braket}
\usepackage{subfigure}
\usepackage[dvipdfmx]{graphicx}
\usepackage{float}
\usepackage{url}
\usepackage{xcolor}
\usepackage{multirow}
\usepackage[colorlinks=true,urlcolor=blue,citecolor=blue,linkcolor=blue]{hyperref}
\usepackage{cleveref}
\usepackage{algorithm}
\usepackage{physics}
\usepackage[noend]{algpseudocode}
\usepackage{comment}
\usepackage{placeins}

\crefname{figure}{Fig.}{Figs.}
\crefname{equation}{Eq.}{Eqs.}
\crefname{section}{Sec.}{Sec.}
\Crefname{figure}{Figure}{Figures}
\Crefname{equation}{Equation}{Equations}
\Crefname{section}{Section}{Sections}

\makeatletter
\newcommand{\subalign}[1]{%
  \vcenter{%
    \Let@ \restore@math@cr \default@tag
    \baselineskip\fontdimen10 \scriptfont\tw@
    \advance\baselineskip\fontdimen12 \scriptfont\tw@
    \lineskip\thr@@\fontdimen8 \scriptfont\thr@@
    \lineskiplimit\lineskip
    \ialign{\hfil$\m@th\scriptstyle##$&$\m@th\scriptstyle{}##$\hfil\crcr
      #1\crcr
    }%
  }%
}
\makeatother

\begin{document}

\title{Randomized benchmarking of a remote CNOT gate via a meter-scale microwave link}

\author{Kentaro Heya}
\email{Kentaro.Heya1@ibm.com}
\affiliation{IBM Quantum, IBM T.J. Watson Research Center, Yorktown Heights, New York 10598, USA}

\author{Timothy Phung}
\affiliation{IBM Quantum, IBM Almaden Research Center, San Jose, CA, 95120, USA}

\author{Moein Malekakhlagh}
\affiliation{IBM Quantum, IBM T.J. Watson Research Center, Yorktown Heights, New York 10598, USA}

\author{Rachel Steiner}
\affiliation{IBM Quantum, IBM T.J. Watson Research Center, Yorktown Heights, New York 10598, USA}

\author{Marco Turchetti}
\affiliation{IBM Quantum, IBM T.J. Watson Research Center, Yorktown Heights, New York 10598, USA}

\author{William Shanks}
\affiliation{IBM Quantum, IBM T.J. Watson Research Center, Yorktown Heights, New York 10598, USA}

\author{John Mamin}
\affiliation{IBM Quantum, IBM Almaden Research Center, San Jose, CA, 95120, USA}

\author{Wen-Sen Lu}
\affiliation{IBM Quantum, IBM T.J. Watson Research Center, Yorktown Heights, New York 10598, USA}

\author{Yadav Prasad Kandel}
\affiliation{IBM Quantum, IBM T.J. Watson Research Center, Yorktown Heights, New York 10598, USA}

\author{Neereja Sundaresan}
\affiliation{IBM Quantum, IBM T.J. Watson Research Center, Yorktown Heights, New York 10598, USA}

\author{Jason Orcutt}
\affiliation{IBM Quantum, IBM T.J. Watson Research Center, Yorktown Heights, New York 10598, USA}

\date{\today}

\begin{abstract}
High-fidelity, meter-scale microwave interconnects between superconducting quantum processor modules are a key technology for extending system size beyond constraints imposed by device manufacturing equipment, yield, and signal delivery.
Although tomographic experiments have been used in previous demonstrations for benchmarking remote state transfer between modules, they do not reliably separate State Preparation and Measurement~(SPAM) error from the error per state transfer.
Recent developments based on randomized benchmarking provide a compatible theory for separating these two errors.
In this work, we present a module-to-module interconnect based on Tunable-Coupling Qubits~(TCQs) and benchmark, in a SPAM-error-tolerant manner enabled by a frame-tracking technique, a remote state transfer fidelity of $98.8$\% across a $60$~cm-long coplanar waveguide~(CPW).
The state transfer is implemented via a superadiabatic transitionless driving method, which suppresses intermediate excitation in the internal modes of the CPW.
We further propose and construct a remote CNOT gate between modules, composed of local CZ gates in each module and remote state transfers, and report a gate fidelity of $93.3$\% using the randomized benchmarking method.
The remote CNOT construction and benchmarking we present provide a way to fully characterize the module-to-module link operation and standardize reporting fidelity, analogous to randomized benchmarking protocols for other quantum gates.
\end{abstract}

\maketitle

Modular design of superconducting quantum processors relaxes wiring complexity and cryogenic cooling power requirements, enables the scaling of the processors beyond constraints imposed by device manufacturing equipment and yield~\cite{bravyi2022future}, and allows the exploration of non-two-dimensional error-correcting schemes~\cite{Campbell_2017}.
For superconducting quantum processors, meter-scale interconnects are a promising technology to connect modules within a fridge, as investigated recently in various proposals~\cite{PhysRevLett.120.200501,chou2018deterministic,kurpiers2018deterministic,axline2018demand,leung2019deterministic,PhysRevLett.125.260502,PhysRevLett.124.240502,PRXQuantum.2.030321,zhong2021deterministic,niu2023low, qiu2025deterministic}.
Although there are a few demonstrations of direct two-qubit gates across interconnects with several hundreds of MHz free spectral range~(FSR) between internal modes~\cite{song2024realizationhighfidelityperfectentangler, Ohfuchi_2024}, they are hard to extend to meter-scale interconnects with several tens of MHz FSR due to frequency collisions~\cite{Malekakhlagh_2020, Brink2018DeviceCF, PRXQuantum.3.020301, doi:10.1126/sciadv.abi6690, Malekakhlagh_Mitigating_2022, kim2022effects, PhysRevApplied.21.024035}.
Two methods to implement indirect two-qubit gates are to (i)~transfer an excitation across the interconnect, or (ii)~generate remote entanglement, such as a Bell state.
These protocols can be broadly categorized as either employing time-symmetric emission and capture of itinerant photons~\cite{Kurpiers_Deterministic_2018, Axline_Demand_2018, Campagne_Deterministic_2018, Zhong_Violating_2019, Bienfait_Phonon_2019, Kannan_Demand_2023, Niu_Low_2023, Qiu_Deterministic_2023}, or using qubit interactions with the standing-wave modes of meter-scale interconnects~\cite{Leung_Deterministic_2019, Zhong_Violating_2019, Chang_Remote_2020, Zhong_Deterministic_2021, Qiu_Deterministic_2023}.
Although the itinerant-photon protocol can extend to longer-distance interconnects, it requires high analog control precision to match the emitter and receiver itinerant photon wave packets to achieve high fidelity.
Therefore, for meter-scale interconnects, protocols based on the standing-wave modes are more suitable due to lower control complexity.
For adiabatic standing-wave protocols, it is also important to suppress excitation in the intermediate interconnect modes to protect against interconnect loss.
To achieve this, we proposed using the SuperAdiabatic Transitionless Driving~(SATD) method~\cite{PhysRevLett.116.230503, zhou2017accelerated} in Ref.~\cite{PhysRevApplied.22.024006}, which is an extension of the STimulated Raman Adiabatic Passage~(STIRAP) method~\cite{Gaubatz_Population_1990, Vitanov_Stimulated_2017, Bergmann_Roadmap_2019, Chang_Remote_2020}, enabling faster state transfer and entangling gates~\cite{PRXQuantum.2.030306, PhysRevApplied.19.034071} with suppressed leakage via shortcuts to adiabaticity.

In this paper, we present a module-to-module microwave interconnect consisting of two TCQs~\cite{PhysRevLett.106.030502, PhysRevB.84.184515, zhang2017suppression, PhysRevLett.106.083601} capacitively coupled to a $60~\mathrm{cm}$-long coplanar waveguide~(CPW), and implement state transfer across the interconnect via SATD.
While tomographic experiments have been used in many previous demonstrations~\cite{PhysRevLett.120.200501,chou2018deterministic,kurpiers2018deterministic,axline2018demand,leung2019deterministic,PhysRevLett.125.260502,PhysRevLett.124.240502,PRXQuantum.2.030321,zhong2021deterministic,niu2023low} to report state transfer fidelity, these do not separate state preparation and measurement~(SPAM) error from the error per state transfer~(EPS).
Therefore, we benchmark an EPS using a SPAM-error-tolerant method as theoretically proposed in Ref.~\cite{helsen2023benchmarking}.
We also implement the remote indirect CNOT gate across the interconnect using multiple state transfers and local CZ gates, and benchmark the error per remote CNOT gate via two-qubit randomized benchmarking.
With these two benchmarks, we propose and demonstrate a robust, standard method for evaluating module-to-module interconnects.

\begin{figure*}
    \centering
	\includegraphics[width=\textwidth]{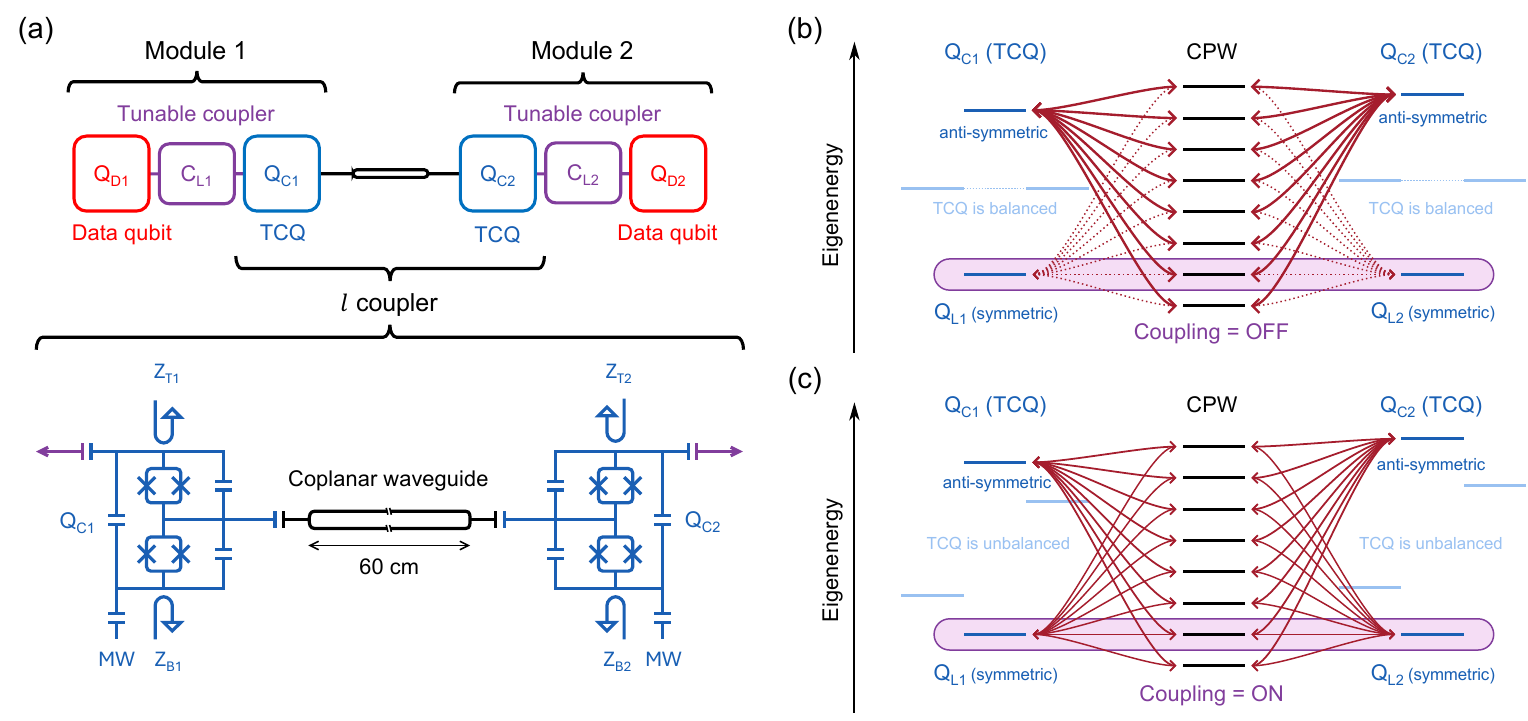}
    \caption{
    (a)
    Circuit schematic consisting of the fixed-frequency data qubits $\mathrm{Q_{D1,2}}$ and the $l$~coupler module coupled via the tunable couplers $\mathrm{C_{L1,2}}$.
    The $l$~coupler module consists of two TCQs~\cite{PhysRevLett.106.030502, PhysRevB.84.184515, zhang2017suppression}, $Q_\mathrm{{C1,2}}$, capacitively coupled to the $60$~cm on-chip CPW.
    The TCQs consist of three electrodes connected in series by two SQUIDs.
    The CPW is coupled to the middle electrodes of the TCQs.
    (b)
    Schematics showing the eigenenergy diagram of the $l$~coupler module when the coupling is switched off.
    Light-blue horizontal lines represent the energies of the top or bottom transmon modes of the TCQs, which will split into blue lines corresponding to the symmetric and anti-symmetric modes.
    By tuning the flux biases of the TCQs, we can independently manipulate the top and bottom mode energies.
    When the top and bottom modes of the TCQs become degenerate with each other (TCQ is balanced), the symmetric modes have voltage nodes at the middle electrode of the TCQ and become decoupled from the CPW modes, shown as black horizontal lines.
    We call the symmetric modes of the TCQs $l$~qubits.
    (c)
    Schematics showing the eigenenergy diagram of the $l$~coupler module when the coupling is switched on.
    When the top and bottom modes of the TCQs are detuned from each other (TCQ is unbalanced), the symmetric modes start to couple to the CPW modes.
    The purple-colored region represents the Hilbert space used for state transfers, where the symmetric modes of the TCQs and the target CPW mode are degenerate with each other and form dark and bright modes, as shown in \cref{eq:bright_dark}.
    During state transfers between $l$~qubits, the flux biases of the TCQs are swept adiabatically between conditions (b) and (c).
    }
    \label{fig:system}
\end{figure*}
A schematic drawing of our module-to-module microwave interconnect is shown in \Cref{fig:system}~(a), which we refer to as an $l$~coupler~\cite{bravyi2022future}.
The $l$~coupler system consists of two TCQs~\cite{PhysRevLett.106.030502, PhysRevB.84.184515, zhang2017suppression}, denoted as $Q_\mathrm{C1,2}$, capacitively coupled to opposite ends of a $60~\mathrm{cm}$-long CPW.
Note that we employed an on-chip CPW for simplicity, but our system is readily extendable to a $1~\mathrm{m}$ coaxial cable interconnect with the same FSR~\cite{martin2024mechanicallyintermixedindiumsuperconductingconnections}.
The long CPW supports multiple resonant modes, separated in frequency by a free spectral range of $\omega_{\mathrm{FSR}}/2\pi = 98~\mathrm{MHz}$.
Each end of the $l$~coupler module is also coupled to fixed-frequency transmon qubits (data qubits)~\cite{PhysRevA.76.042319}, referred to as $Q_\mathrm{D1,2}$, via tunable couplers~\cite{PhysRevApplied.6.064007, Yan_Tunable_2018, Foxen_Demonstrating_2020, Collodo_Implementation_2020, Stehlik_Tunable_2021, Sung_Realization_2021, PhysRevApplied.16.024050, zajac2021spectator, Li_Realization_2024}, denoted as $\mathrm{C_{L1,2}}$, respectively.
Each qubit is coupled to microwave lines for independent qubit control and readout via a multiplexed Purcell filter~\cite{PhysRevLett.112.190504,PhysRevA.92.012325}.
We read out the qubit states using standard dispersive measurements~\cite{PhysRevA.69.062320,PhysRevLett.95.060501} with the aid of a traveling-wave parametric amplifier~\cite{PhysRevB.87.144301}.
Each TCQ has two flux lines, which permit independent manipulation of the detuning and coupling strength between the TCQ and the CPW modes.
The tunable couplers $\mathrm{C_{L1,2}}$ are used to implement local CZ gates between the data qubits and the TCQs.
The system parameters are summarized in \cref{Tab:device_parameters}.

\begin{figure*}
    \centering
	\includegraphics[width=\textwidth]{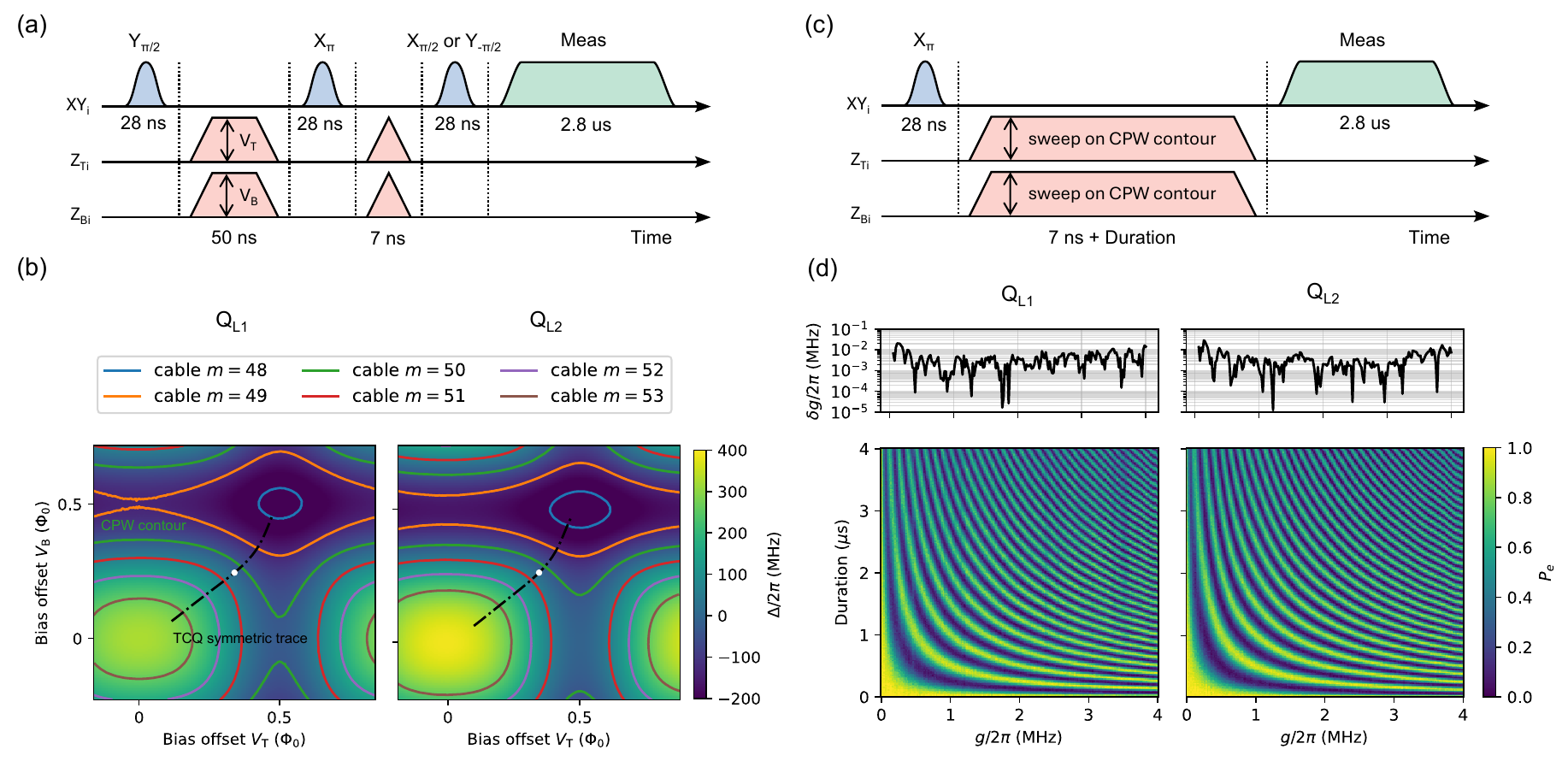}
    \caption{
    (a)
    Pulse sequence for the fixed-duration XY Ramsey experiments to estimate the $l$~qubit eigenfrequencies while independently sweeping the top and bottom flux pulse amplitudes $V_\mathrm{T,B}$.
    The second flux pulse and the echo microwave pulse are used to cancel out the $l$~qubit eigenfrequency shifts at the edges of the flux pulses.
    (b)
    Experimental results of (a).
    Contour lines correspond to the CPW mode frequencies.
    Green contour lines represent the target CPW mode used for state transfers.
    Black dotted lines represent the trace where the TCQs become symmetric.
    White dots indicate the idle bias condition.
    (c)
    Pulse sequence for the vacuum Rabi oscillation experiment to evaluate the coupling strength between the $l$~qubits and the target CPW mode ($m=50$).
    (d)
    Experimental results of (c).
    We sweep the top and bottom flux biases along the green contour lines in \cref{fig:control_accuracy}~(a) to vary the couplings linearly.
    The top panels show the calibration error of the coupling strengths.
    }
    \label{fig:control_accuracy}
\end{figure*}
As shown in \cref{fig:system}~(a), each TCQ consists of two frequency-tunable transmon qubits connected galvanically to each other and sharing a common charge island.
We can consider that each TCQ behaves like two individual transmons when the difference in the two transmon frequencies is much larger than the coupling strength between them.
In this case, one can think of the device as having two individual modes of excitation, which we refer to as the top and bottom modes.
Both modes will have finite coupling to the CPW connecting the TCQs.
However, when the two individual mode frequencies are in resonance, the two modes become hybridized, and symmetric and anti-symmetric modes emerge, as shown in \cref{fig:system}~(b) and (c).
Both modes have distinct electric dipole moments and thus couple to the CPW differently, with the symmetric mode having zero coupling to the CPW.
Details about the physics and design principles of TCQs are shown in the Supplemental Material.

While the TCQ has a multimode structure, we shall use only the symmetric modes shown in \cref{fig:system}~(b) and (c) for operations in the $l$~coupler, henceforth referred to as $l$~qubits $Q_{\mathrm{L1,2}}$, corresponding to the TCQs $Q_\mathrm{C1,2}$, respectively.
We initially store data that we wish to transfer across the CPW in the $l$~qubit, which is dark to the CPW modes.
By applying flux pulses, we can independently control the coupling and detuning between $l$~qubits and the CPW modes.
In addition, this can be done adiabatically with respect to the coupling between the $l$~qubits and the anti-symmetric modes.

\Cref{fig:control_accuracy} shows the pulse sequences and experimental results on the $l$~coupler module, demonstrating the controllability of the detuning and coupling to the target CPW mode.
Color plots in \cref{fig:control_accuracy}~(b) show the $l$~qubit eigenfrequencies depending on the top and bottom flux biases, which are reconstructed from the rotational phase in the fixed-duration Ramsey experiments with a $36$~ns delay time, shown in \cref{fig:control_accuracy}~(a).
The contours in \cref{fig:control_accuracy}~(b) correspond to CPW mode frequencies.
The white dot corresponds to the flux biases at the idle condition.
Color plots in \cref{fig:control_accuracy}~(d) show the vacuum Rabi oscillation between the $l$~qubits and the target CPW mode ($m=50$), where we make the $l$~qubits degenerate with the target CPW mode while sweeping the coupling strength linearly from $0$ to $4$~MHz, as shown in \cref{fig:control_accuracy}~(b).
The top plots of \cref{fig:control_accuracy}~(d) show the calibration error in the coupling strengths, which is approximately $10~\mathrm{kHz}$.

Note that the maximum coupling strength is limited by the capacitive coupling between the TCQs and the CPW and is not a fundamental limitation of the TCQ-based interconnects.
Since our methodology of state transfer allows the ratio of the maximum coupling strength to the CPW FSR to be up to $0.15$~\cite{PhysRevApplied.22.024006, Chang_Remote_2020}, we can expect to achieve higher state transfer fidelity by increasing the maximum coupling while maintaining the current CPW length.

\begin{figure*}
    \centering
	\includegraphics[width=0.9\textwidth]{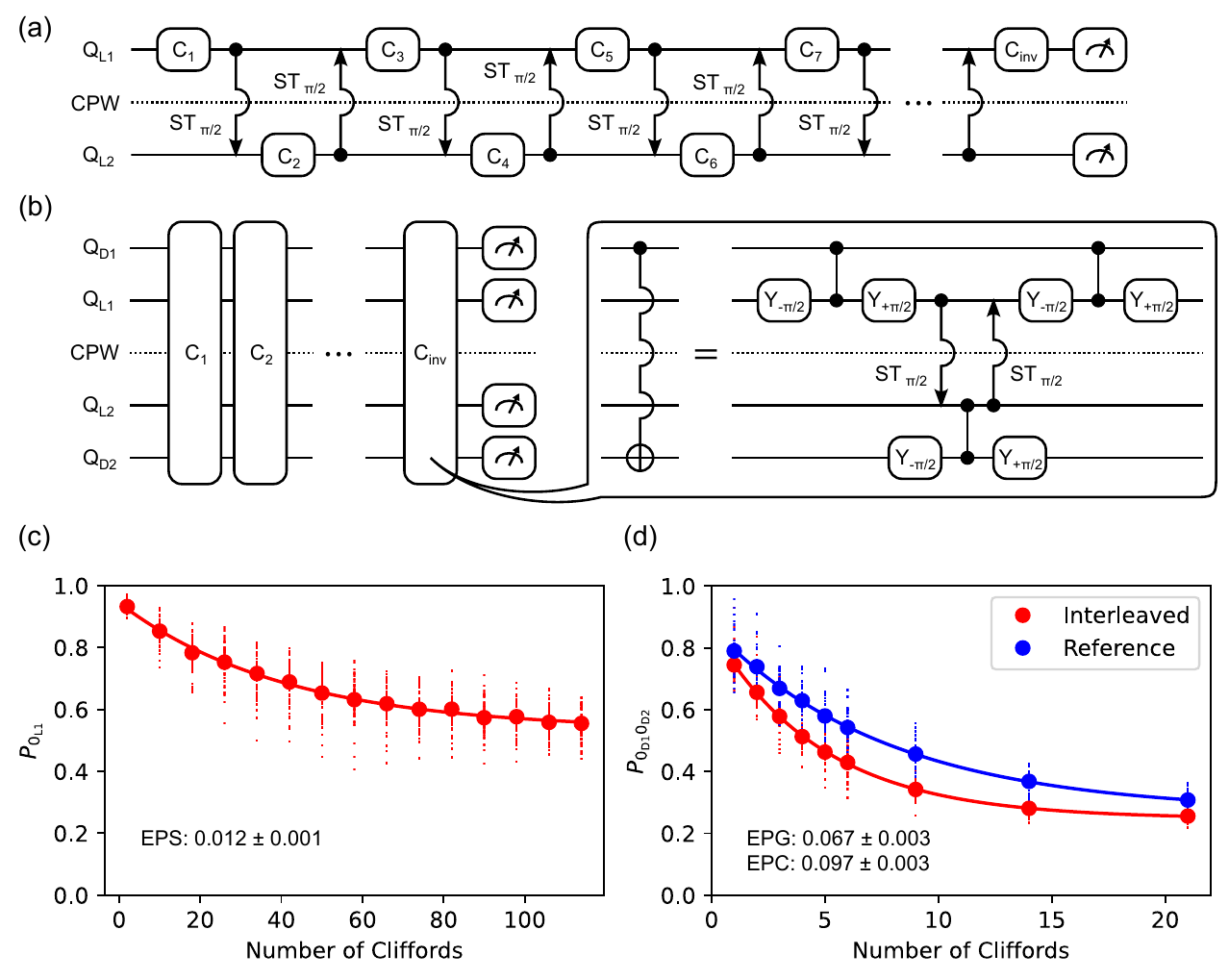}
    \caption{
    (a)
    Quantum circuit for the network benchmarking~\cite{helsen2023benchmarking}.
    $\mathrm{C_i}$ represents a random single-qubit Clifford gate.
    $\mathrm{C_{inv}}$ represents the inverse Clifford operation to make the total sequence equal to the identity operation.
    Arrows $\mathrm{ST_{\pi/2}}$ in the circuit represent directional state transfer between the $l$~qubits via the CPW.
    Note that, although the $l$~qubit population is transferred in both directions in SATD-based state transfer, only in one direction is it transferred with higher fidelity via the dark state.
    (b)
    Quantum circuit for the two-qubit randomized benchmarking of the remote CNOT gate.
    $\mathrm{C_i}$ represents a random two-qubit Clifford gate between the data qubits.
    $\mathrm{C_{inv}}$ represents the inverse Clifford operation to make the total sequence equal to the identity operation.
    Each remote CNOT gate consists of three local CZ gates at each module and two directional state transfers between the modules.
    $\mathrm{Y_{\pm\pi/2}}$ represent the $\pm\pi/2$ rotation on the Pauli Y axis, respectively.
    (c)
    Network benchmarking on the $l$~qubits: probability of measuring $Q_\mathrm{L1}$ in the $\ket{0}$ state as a function of the number of Cliffords.
    We measure an error per state transfer~(EPS) of 0.012.
    (d)
    Two-qubit randomized benchmarking on the data qubits for the remote CNOT gate from $Q_\mathrm{D1}$ to $Q_\mathrm{D2}$: probability of measuring the data qubits $Q_\mathrm{D1,2}$ in the $\ket{00}$ state as a function of the number of Cliffords.
    EPC and EPG represent the error per Clifford and the error per gate, respectively.
    We measure an EPG of 0.067.
    }
    \label{fig:rb}
\end{figure*}

State transfer between the $l$~qubits is performed using the superadiabatic transitionless driving~(SATD) method~\cite{PhysRevApplied.22.024006, PhysRevLett.116.230503, zhou2017accelerated}.
Under the SATD method, we first prepare one of the $l$~qubits in the excited state and the other in the ground state, referred to as the emitter and receiver qubits, respectively.
We linearly modulate the $l$~qubit frequencies from their idle positions and make them degenerate with the target CPW mode, while keeping the coupling between the $l$~qubits and the target CPW mode ($m=50$) at zero.
While the frequencies are degenerate, we turn on the coupling between the $l$~qubits and the target CPW mode according to:
\begin{align} \label{eq:coupling_satd}
g_\mathrm{e} (t)
&=g\qty[\sin\theta(t)+\frac{\ddot{\theta}(t)\cos\theta(t)}{g^2 + \dot{\theta}^2(t)}]\;, \\
g_\mathrm{r} (t)
&=g\qty[\cos\theta(t)-\frac{\ddot{\theta}(t)\sin\theta(t)}{g^2 + \dot{\theta}^2(t)}]\;, \\
\theta(t)&=\frac{\pi}{2}\qty{6\qty(\frac{t}{T})^5 - 15\qty(\frac{t}{T})^4 + 10\qty(\frac{t}{T})^3}\;,
\end{align}
where $g_\mathrm{e,r}(t)$ represent the coupling strengths between the emitter and receiver qubits and the target CPW mode, respectively.
Derivations of \cref{eq:coupling_satd} are provided in the Supplemental Materials.
In the following experiments, we use a maximum coupling strength of $g/2\pi=3.5~\mathrm{MHz}$ and an execution time of the protocol $T=135~\mathrm{ns}$.
While sweeping the couplings, the composite Hilbert space spanned by the $l$~qubits and the target CPW mode has the following \textit{instantaneous} eigenstates:
\begin{align}\label{eq:bright_dark}
\ket{D(t)}\equiv
&\cos\theta(t)\ket{100} - \sin\theta(t)\ket{001}\;,\\
\ket{B_{\pm}(t)}\equiv
&\frac{1}{\sqrt{2}}\left(\sin\theta(t)\ket{100}\pm\ket{010} \right. \nonumber\\
& \left. -\cos\theta(t)\ket{001}\right)\;,
\end{align}
where the state $\ket{ijk}$ represents the $i$, $j$, and $k$th excitations in the emitter qubit, target CPW mode, and receiver qubit, respectively.
Here, $\ket{D(t)}$ and $\ket{B_\pm(t)}$ are called the dark and bright eigenstates, respectively.
The dark eigenstate, having no overlap with the intermediate (possibly lossy) interconnect state, therefore allows for adiabatic quantum state transfer by arbitrarily evolving the mixing angle $\theta(t)$~\cite{Gaubatz_Population_1990, Vitanov_Stimulated_2017, Bergmann_Roadmap_2019}.
The main advantage of SATD is the suppression of diabatic leakage from the dark state to the bright states using the counter-adiabatic corrections~\cite{PhysRevApplied.22.024006}.
Finally, we turn off the couplings between both $l$~qubits and the target CPW mode and linearly move both $l$~qubit frequencies back to their idle frequencies.
The overall SATD sequence described above takes $206$~ns on this device.
More details on experiments regarding leakage suppression using SATD can be found in the Supplemental Material.

We implement directional state transfer using the SATD method in both directions and evaluate the error per state transfer~(EPS) via the network benchmarking~(NB) method of Ref.~\cite{helsen2023benchmarking}.
The protocol and the results of NB are shown in \cref{fig:rb}~(a) and (c), where we perform single-qubit randomized benchmarking with alternating $l$~qubits swapped by directional state transfers.
In our NB implementation, the number of Clifford gates is always chosen to be an even number to ensure that the final population returns to $Q_\mathrm{L1}$.
The exponential curve fit of the $Q_\mathrm{L1}$ ground state population at the end of the NB circuit yields the base $\gamma_\mathrm{NB}$.
The EPS is converted from the base as $\varepsilon_\mathrm{ST}=(1-\gamma)/2$, resulting in $0.012\pm 0.001$ in our experiment.
Note that the NB circuit and the following two-qubit RB experiments, which consist of multiple state transfers, require frame tracking~\cite{PhysRevApplied.21.024018, PRXQuantum.5.020338} to compensate for unintended phase rotations on the $l$~qubits due to qubit frequency differences before and after each state transfer.
More details about the frame tracking can be found in the Supplemental Material.
A detailed error budget analysis of the SATD-based state transfer is provided in the Supplemental Material, where the largest error source is energy decay and decoherence, accounting for approximately $2/3$ of the total error.
As shown in the Supplemental Material, even though we always use an even number of state transfers, we observe an increase in the population of $Q_\mathrm{L2}$ with sequence length, which we hypothesize originates from leakage errors during state transfer.
The leakage rate is also estimated from the exponential fit of the final population of $Q_\mathrm{L2}$ to be $0.011\pm0.001$.

The control line to the $i$th tunable coupler $\mathrm{C_{Li}}$ is used to implement local CZ gates between the data qubit $Q_\mathrm{Di}$ and $l$~qubit $Q_\mathrm{Li}$.
The gate lengths of the CZ gates between $Q_\mathrm{L1}$-$Q_\mathrm{D1}$ and $Q_\mathrm{L2}$-$Q_\mathrm{D2}$ are $135$~ns and $100$~ns, respectively.
The error per gate~(EPG) of the calibrated CZ gates is evaluated by interleaved two-qubit randomized benchmarking~(TQRB)~\cite{PhysRevLett.109.080505} as $0.0093\pm0.0003$ and $0.0054\pm0.0003$ for the qubit pairs $Q_{\mathrm{L1}}$-$Q_{\mathrm{D1}}$ and $Q_{\mathrm{L2}}$-$Q_{\mathrm{D2}}$, respectively.

The remote CNOT gate between the data qubits via the $l$~coupler module is implemented through composite pulse sequences including three local CZ gates at each module and two directional state transfers between the modules, as shown in \cref{fig:rb}~(b).
The total gate length of the remote CNOT gate is $957$~ns.
\Cref{fig:rb}~(d) shows the probability of measuring the data qubits $Q_\mathrm{D1,2}$ in the $\ket{00}$ state as a function of the number of Cliffords.
We measure an EPG of $0.067\pm0.003$.
We note that when the noise of the underlying gates covering the remote CNOT gate is approximated by a depolarizing channel inferred from the measured EPG, the resulting EPG of the remote CNOT gate is $0.063$.
Similar to the NB result, we observe an increase in the population not only in the data qubits but also in the $l$~qubits due to leakage during state transfer.
As shown in the Supplemental Material, the leakage rates on the $Q_{\mathrm{L1,2}}$ qubits are evaluated as $0.050\pm0.003$ and $0.040\pm0.005$, respectively.

In this work, we develop and benchmark a module-to-module microwave interconnect to scale up superconducting quantum processors.
Our microwave interconnect consists of two tunable-coupling qubits~(TCQs) capacitively coupled to opposite ends of a $60$~cm CPW, which are also coupled to fixed-frequency data qubits via tunable couplers.
The state transfer between the symmetric modes of the TCQs is performed via the SATD method, allowing faster adiabatic state transfer via shortcuts to adiabaticity.
We also introduce the frame tracking technique to circumvent residual phase rotations, which appear on the qubits before and after each state transfer due to frame rotation at each qubit eigenfrequency.
Our frame tracking enables the implementation of SPAM error-tolerant benchmarking experiments such as network benchmarking and randomized benchmarking.
In the benchmarking experiments, we demonstrate state transfer between the TCQs with an EPS of $0.011$ and the remote CNOT gate between the data qubits with an EPG of $0.067$.
Our work proposes and demonstrates a standard method for evaluating module-to-module interconnects.

\begin{acknowledgments}
\emph{Acknowledgement}---We appreciate helpful discussions with the IBM Quantum members Luke Govia, Ken Xuan Wei, Ali Javadi-Abhari, Aaron Finck, David Mckay, Holger Haas, George Stehlik, David Zajac, Seth Merkel, and Muir Kumph.
\end{acknowledgments}

\bibliography{main.bib}

\clearpage
\widetext
\begin{center}
\textbf{\large Supplementary Information: Randomized benchmarking of a remote CNOT gate via a meter-scale microwave link}
\end{center}

\setcounter{equation}{0}
\setcounter{figure}{0}
\setcounter{table}{0}
\setcounter{page}{1}
\makeatletter
\renewcommand{\theequation}{S\arabic{equation}}
\renewcommand{\thefigure}{S\arabic{figure}}
\renewcommand{\bibnumfmt}[1]{[S#1]}
\renewcommand{\citenumfont}[1]{S#1}

\FloatBarrier

\section{System parameters}
The parameters of the fixed-frequency data qubits, the $l$~qubits, and the CPW modes are listed in \cref{Tab:device_parameters}.
The idle frequencies of the anti-symmetric modes of the TCQs $Q_\mathrm{C1,2}$ are determined from spectroscopy experiments as $5.463$ and $5.475$~GHz, respectively.
The intrinsic lifetimes of the CPW modes were measured using the following procedure.
We first prepare the $\ket{1}$ state in $Q_\mathrm{L1}$ and transfer it to the target CPW modes using simultaneous shaped flux pulses on $Z_\mathrm{T,B}$.
We then wait for a finite amount of time and transfer the population back from the target CPW modes to $Q_\mathrm{L1}$.
Finally, we observe the population decay in $Q_\mathrm{L1}$ while sweeping the delay time.

\begin{table*}
\caption{Device parameters.}
\begin{tabular}{p{9cm}cccc}
\hline \hline
Qubit parameters                                        & $Q_\mathrm{L1}$ & $Q_\mathrm{L2}$ & $Q_\mathrm{D1}$ & $Q_\mathrm{D2}$ \\
\hline
Qubit maximum frequency $\omega_{ge}^{max}/2\pi$~(GHz)  &           5.243 &           5.310 &               - &               - \\
Qubit minimum frequency $\omega_{ge}^{min}/2\pi$~(GHz)  &           4.681 &           4.665 &               - &               - \\
Qubit idle frequency $\omega_{ge}^{idle}/2\pi$~(GHz)    &           4.933 &           4.929 &               - &               - \\
Qubit anharmonicity $\alpha/2\pi$~(MHz)                 &            -134 &            -134 &               - &               - \\
Qubit intrinsic lifetime,
$T_{1,int}$~($\mathrm{\mu s}$)                          &             113 &              76 &              92 &             141 \\
Qubit Ramsey dephasing time,
$T_{2,ramsey}$~($\mathrm{\mu s}$)                       &              18 &              22 &              38 &              81 \\
Readout resonator frequency, $\omega_r/2\pi$~(GHz)      &           7.089 &           6.866 &           6.915 &           7.033 \\
Readout assignment fidelity, $F_{\mathrm{readout}}$     &           0.959 &           0.957 &           0.967 &           0.976 \\
Initial thermal population, $P_{e}$~(\%)                &             1.9 &             2.0 &             1.9 &             1.3 \\
Error per single-qubit Clifford gates,
$\varepsilon_{1q}$~($10^{-4}$)                          &        $7.8(2)$ &        $8.0(3)$ &        $4.7(1)$ &        $2.5(1)$ \\
\hline \hline
\  & \  & \  & \ \\
\hline \hline
CPW parameters                                          &          $m=49$ &          $m=50$ &          $m=51$ &          $m=52$ \\
\hline
CPW mode frequency $\omega_r/2\pi$~(GHz)                &           4.783 &           4.881 &           4.980 &           5.078 \\
CPW mode intrinsic lifetime,
$T_{1,int}$~($\mathrm{\mu s}$)                          &            5.15 &            5.23 &            5.13 &            4.53 \\
\hline \hline
\  & \  & \  & \ \\
\hline \hline
Coupler parameters                                      & $Q_\mathrm{D1}-Q_\mathrm{L1}$  & $Q_\mathrm{L1}-Q_\mathrm{L2}$ & $Q_\mathrm{L2}-Q_\mathrm{D2}$ \\
\hline
CZ gate error per gate,
$\varepsilon_{cz}$~($10^{-2}$)                          &                      $0.93(3)$ &                             - &                     $0.54(3)$ \\
\hline \hline
\end{tabular}
\label{Tab:device_parameters}
\end{table*}

\section{Physics and design principle of TCQs}
\begin{figure*}
    \centering
	\includegraphics[width=6cm]{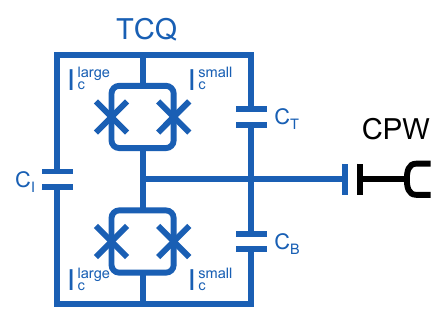}
    \caption{
    Partial circuit diagram of the $l$-coupler module consisting of a TCQ and a CPW coupled to each other capacitively.
    }
    \label{fig:tcq_phys}
\end{figure*}

Tunable-coupling qubits~(TCQs) consist of two frequency-tunable transmon qubits connected galvanically to each other and sharing a common charge island.
The circuit diagram is shown in \cref{fig:tcq_phys}.
The charge-operator-basis Hamiltonian of a TCQ is written as:
\begin{align}
H=\sum_{i\in[T,B]} \qty{4E_{Ci}(n_i - n_{gi})^2 -E_{Ji}\cos(\phi_i)} + 4E_I n_1 n_2 \;,
\end{align}
with charge energy $E_{Ci}=e^2/2C_i$, Josephson energy $E_{Ji}$, charge operator $n_i$, phase operator $\phi_i$, and dimensionless offset charge $n_{gi}$ for the top ($i=T$) or bottom ($i=B$) transmon mode, respectively.
The two transmon modes are coupled via an interaction energy $E_I=e^2/2C_I$.
Note that both $E_{Ji}$ derive from the $i$th SQUID in the TCQ and are tunable using the corresponding flux lines $Z_{i}$.
Under a Duffing-oscillator approximation, the Hamiltonian with a rotating-wave approximation can be written as:
\begin{align}
\frac{H}{\hbar}=\sum_{i\in[T,B]} \qty\big{\omega_i \hat{a}^\dagger_i \hat{a}_i + \frac{\alpha_i}{2}\hat{a}^\dagger_i\hat{a}^\dagger_i\hat{a}_i\hat{a}_i} + J_\mathrm{TB}(\hat{a}^\dagger_T\hat{a}_B + \hat{a}_T \hat{a}^\dagger_B) \;,
\end{align}
with $\omega_{i}/\hbar=\sqrt{8E_{Ci}E_{Ji}}-E_{Ci}$, $\alpha_{i}/\hbar=-E_{Ci}$, and $J_\mathrm{TB}/\hbar=E_I (E_{JT}E_{JB}/E_{CT}E_{CB})^{1/4}$.
We employ symmetric electrodes $C_T=C_B$ in TCQs, which results in the same anharmonicities $\alpha_{T,B}=\bar{\alpha}$ for the transmon modes.
At idle bias, we allocate the two transmon modes in TCQs symmetrically as $\omega_T = \omega_B = \bar{\omega}$.
Therefore, the diagonalized Hamiltonian is written as:
\begin{align}
\frac{H}{\hbar}=\sum_{i\in[+,-]} \qty\big{\omega_i \tilde{\hat{a}}^\dagger_i \tilde{\hat{a}}_i + \frac{\alpha_i}{2}\hat{a}^\dagger_i\hat{a}^\dagger_i\hat{a}_i\hat{a}_i} + \xi \hat{a}^\dagger_+\hat{a}_+ \hat{a}^\dagger_-\hat{a}_- \;,
\end{align}
with $\omega_\pm=\bar{\omega}\pm J_\mathrm{TB}$, $\alpha_\pm=(\alpha\pm\sqrt{\alpha^2+16J_\mathrm{TB}^2})/2\mp2J_\mathrm{TB}$, and $\xi=\alpha/2$.
The index $i$ corresponds to the symmetric ($i=-$) or anti-symmetric ($i=+$) hybridized modes, respectively.
To achieve the target parameters in \cref{Tab:device_parameters}, we design the circuit parameters of TCQs as $C_T=C_B=73.4~\mathrm{fF}$ and $C_I=8.4~\mathrm{fF}$.
Each TCQ has identically designed asymmetric DC-SQUIDs consisting of large and small Josephson junctions with critical currents $I_c^\mathrm{large}=32.5~\mathrm{nA}$ and $I_c^\mathrm{small}=5.4~\mathrm{nA}$, respectively.

\section{Controls of detuning and coupling between TCQs and a CPW mode}
Let us assume a TCQ capacitively coupled to a CPW, where the Hamiltonian in the single-excitation subspace is written as:
\begin{align}
\frac{H_1}{\hbar}
=\sum_{i=[\mathrm{T},\mathrm{B},\mathrm{c}]}\omega_i\ket{i}\bra{i}
+ J_\mathrm{TB}\qty(\ket{T}\bra{B}+\ket{B}\bra{T})
+ \sum_{i=[\mathrm{T},\mathrm{B}]} g_\mathrm{ic}\qty(\ket{i}\bra{c}+\ket{c}\bra{i})
\end{align}
with $\omega_i$ being the eigenfrequency of the $i$th mode, $J_\mathrm{TB}$ the coupling between the top and bottom modes of the TCQ, and $g_{ic}$ the coupling between the TCQ modes and the CPW, where the subscripts represent the top ($i=\mathrm{T}$) and bottom ($i=\mathrm{B}$) modes of the TCQ, and a CPW mode $i=c$.
Diagonalizing the $2\times2$ submatrix in the upper left transforms the Hamiltonian as:
\begin{align}
\frac{H'_1}{\hbar}=
\begin{pmatrix}
\frac{\omega_\mathrm{T}+\omega_\mathrm{B}-\sqrt{(\omega_\mathrm{T}-\omega_\mathrm{B})^2+4J_\mathrm{TB}^2}}{2} & 0 & g_{\mathrm{T}c}\cos\phi_\mathrm{TB}-g_{\mathrm{B}c}\sin\phi_\mathrm{TB} \\
0 & \frac{\omega_\mathrm{T}+\omega_\mathrm{B}+\sqrt{(\omega_\mathrm{T}-\omega_\mathrm{B})^2+4J_\mathrm{TB}^2}}{2} & g_{\mathrm{B}c}\cos\phi_\mathrm{TB}+g_{\mathrm{T}c}\sin\phi_\mathrm{TB} \\
g_{\mathrm{T}c}\cos\phi_\mathrm{TB}-g_{\mathrm{B}c}\sin\phi_\mathrm{TB} & g_{\mathrm{B}c}\cos\phi_\mathrm{TB}+g_{\mathrm{T}c}\sin\phi_\mathrm{TB} & \omega_c
\end{pmatrix}\;,
\end{align}
with the mixing angle $\phi_\mathrm{TB}=\arctan\qty(2J_\mathrm{TB}/(\omega_\mathrm{T}-\omega_\mathrm{B}))/2$.
Because the TCQs have symmetric electrodes and couple to the CPW only via the center electrode, we can assume the symmetric coupling condition $g_{\mathrm{T}c}=g_{\mathrm{B}c}=g_{c}$.
In our design, the coupling rate is set to $g_{c}/2\pi=6.8~\mathrm{MHz}$.
Therefore, the coupling $g_{sc}$ and detuning $\Delta_{sc}$ between the TCQ symmetric eigenmode ($l$~qubit) and a CPW mode are written as:
\begin{align}
\label{eq:tunable_coupling_tcq}
g_{sc} &= \sqrt{2} g_c \cos\qty(\phi_\mathrm{TB} + \frac{\pi}{4}), \\
\Delta_{sc} &= \frac{\omega_\mathrm{T}+\omega_\mathrm{B}-\sqrt{(\omega_\mathrm{T}-\omega_\mathrm{B})^2+4J_\mathrm{TB}^2}}{2} - \omega_c\;.
\end{align}
In this manuscript, we treat the TCQ symmetric eigenmodes as $l$~qubits.
To perform a state transfer between $l$~qubits and a CPW mode, we manipulate the two flux-bias knobs of each TCQ to sweep the coupling between the $l$~qubits and the CPW mode as in \cref{eq:coupling_satd}, while stabilizing the detuning between them at $0$.
Because the analytical expressions of the flux biases required to achieve these conditions become complicated, we employ a calibration strategy based on experiments described in \cref{fig:control_accuracy}.
We first characterize the $i$th $l$~qubit frequency as a function of the two flux pulse amplitudes of the corresponding TCQ as $f_i(\phi_\mathrm{Ti}, \phi_\mathrm{Bi})$ using a bivariate spline approximation, where $\phi_\mathrm{Ti}$ and $\phi_\mathrm{Bi}$ represent the flux pulse amplitudes applied to flux lines $Z_\mathrm{Ti}$ and $Z_\mathrm{Bi}$, respectively.
From these frequency functions, we can calculate contour lines of flux pulse amplitudes that make the $l$~qubits degenerate with the target CPW mode as $\phi_\mathrm{Ti}=h_\mathrm{Ti}(\beta_i)$ and $\phi_\mathrm{Bi}=h_\mathrm{Bi}(\beta_i)$, with the contour functions $h_\mathrm{Ti}$ and $h_\mathrm{Bi}$ calculated by interpolation, and the contour parameter $\beta_i$ corresponding to a coordinate along the contour line.
We next observe vacuum Rabi oscillations between the $i$th $l$~qubit and the target CPW mode while sweeping the contour parameter $\beta_i$, where the vacuum Rabi oscillation frequency is interpolated with a fitting function $\nu_i(\beta_i)$.
In SATD-based state transfers, $l$~qubits must become degenerate with the target CPW mode while the coupling strengths are swept as in \cref{eq:coupling_satd}.
Therefore, the control flux pulse shapes are calculated as:
\begin{align}
\phi_{Te}(t)&=h_{Te}\qty(\nu_e^{-1}\qty(g_e(t)))\;, \\
\phi_{Be}(t)&=h_{Be}\qty(\nu_e^{-1}\qty(g_e(t)))\;, \\
\phi_{Tr}(t)&=h_{Tr}\qty(\nu_r^{-1}\qty(g_r(t)))\;, \\
\phi_{Br}(t)&=h_{Br}\qty(\nu_r^{-1}\qty(g_r(t)))\;,
\end{align}
where $i=e,r$ represents the emitter and receiver $l$~qubit, respectively.

\section{Derivation of coupling formulas for STIRAP and SATD}
In this section, we review the derivation of STIRAP and SATD.
A similar derivation can also be found in our previous theoretical proposal~\cite{PhysRevApplied.22.024006}.

Consider the following single-excitation manifold Hamiltonian for a resonant Lambda system with tunable interaction rates:
\begin{align}
H_\Lambda (t)=
\begin{pmatrix}
0 & g(t)\sin\theta(t) & 0 \\
g(t)\sin\theta(t) & 0 & g(t)\cos\theta(t) \\
0 & g(t)\cos\theta(t) & 0
\end{pmatrix},
\end{align}
where $g(t)$ is a time-dependent coupling and $\theta(t)$ is the mixing angle.
The modes corresponding to each row of the Hamiltonian are an emitter qubit, a lossy interconnect, and a receiver qubit, respectively.

The instantaneous Hamiltonian can be approximately diagonalized as:
\begin{align}
H'_\Lambda (t)
&=
\mathcal{U}(H_\Lambda(t)) \\
&=
\begin{pmatrix}
+g(t) & 0 & 0 \\
0 & 0 & 0 \\
0 & 0 & -g(t)
\end{pmatrix}
+
\begin{pmatrix}
0 & +i\dot{\theta}(t) & 0 \\
-i\dot{\theta}(t) & 0 & -i\dot{\theta}(t) \\
0 & +i\dot{\theta}(t) & 0
\end{pmatrix},
\label{eq:instant_frame}
\end{align}
via the instantaneous unitary transformation
\begin{align}
\mathcal{U}(t)(H_\Lambda(t)) &= U(t) H_\Lambda (t) U^\dagger (t) + i\dot{U}(t) U^\dagger (t), \\
U(t) &=
\begin{pmatrix}
\frac{\sin\theta(t)}{\sqrt{2}} & +\frac{1}{\sqrt{2}} & \frac{\cos\theta(t)}{\sqrt{2}} \\
\cos\theta(t) & 0 & -\sin\theta(t) \\
\frac{\sin\theta(t)}{\sqrt{2}} & -\frac{1}{\sqrt{2}} & \frac{\cos\theta(t)}{\sqrt{2}}
\end{pmatrix}.
\end{align}

The first and second terms in \cref{eq:instant_frame} represent the adiabatic term and the non-adiabatic transition term, respectively.
STIRAP operates under the adiabatic condition, in which the rate of change of the mixing angle is much slower than the maximum coupling strength.

Under this condition, the instantaneous eigenstates of the Hamiltonian are:
\begin{align}
\ket{D(t)} &\equiv \cos\theta(t)\ket{100} - \sin\theta(t)\ket{001},\\
\ket{B_{\pm}(t)} &\equiv \frac{1}{\sqrt{2}}\left(\sin\theta(t)\ket{100}\pm\ket{010} -\cos\theta(t)\ket{001}\right),
\end{align}
where $\ket{D(t)}$ and $\ket{B_{\pm}(t)}$ are called the dark and bright states, respectively.

In STIRAP, we typically choose a fixed coupling $g(t)=g$ and transfer an initial quantum state in the emitter qubit to the receiver qubit by linearly sweeping the mixing angle from $0$ to $\pi/2$ through the dark state $\ket{D(t)}$ without exciting the lossy interconnect.  
While typical state transfers using exchange interactions are symmetric with respect to the qubits and provide bidirectional transfer, STIRAP achieves unidirectional state transfer via the dark state, which is protected from losses in the interconnect.  
Therefore, we refer to this as a directional state transfer.

SATD is an extension of STIRAP designed to mitigate non-adiabatic leakage from the dark state to the bright states.
Revisiting \cref{eq:instant_frame}, the diagonalized instantaneous Hamiltonian can be written as:
\begin{align}
H'_\Lambda (t) = g(t) M_z + \dot{\theta}(t) M_y \;,
\end{align}
where $M_k$ ($k=x,y,z$) are spin-1 operators.

We introduce a Hamiltonian for the SATD-based state transfer, $H_{\Lambda+c}(t)$, with a counter-adiabatic correction $H_\mathrm{c}(t)$:
\begin{align}
H_{\Lambda+c}(t) &= H_\Lambda(t) + H_\mathrm{c}(t), \\
H_\mathrm{c}(t) &= U^\dagger(t) \qty[\lambda_x(t) M_x] U(t),
\end{align}
where $\lambda_x(t)$ is a time-dependent amplitude and $U(t)$ is the instantaneous unitary defined previously.
We also define a dressed-frame unitary transformation $V(t)$ for an arbitrary operator $X$:
\begin{align}
\mathcal{V}(t)(X) &= V(t) X V^\dagger(t) - i \dot{V}(t) V^\dagger(t), \\
V(t) &= \exp\qty{i \phi_x(t) M_x}, \label{eq:satd_frame_change}
\end{align}
with a time-dependent angle $\phi_x(t)$.

The SATD Hamiltonian in the dressed frame is diagonalized as:
\begin{align}
H''_{\Lambda+c}(t)
&= \mathcal{V}(t) \qty(\mathcal{U}(t)(H_{\Lambda+c}(t))) \nonumber \\
&=
\qty[g(t)\cos\phi_x(t)-\dot{\theta}(t)\sin\phi_x(t)] M_z
+ \qty[g(t)\sin\phi_x(t) + \dot{\theta}(t)\cos\phi_x(t)] M_y
+ \qty[\lambda_x(t) - \dot{\phi}_x(t)] M_x.
\end{align}

To eliminate non-diagonal terms in the dressed-frame Hamiltonian, we require:
\begin{align}
\phi_x(t) &= \arctan\qty(-\frac{\dot{\theta}(t)}{g(t)}), \\
\lambda_x(t) &= \dot{\phi}_x(t),
\end{align}
with an explicit solution for $\lambda_x(t)$:
\begin{align}
\lambda_x(t) = -\frac{g \ddot{\theta}(t)}{g^2 + \dot{\theta}^2(t)}.
\end{align}

Transforming back to the lab frame, the SATD Hamiltonian becomes:
\begin{align}
H_\Lambda(t) + H_\mathrm{ctrl}(t) =
\begin{pmatrix}
0 & g\qty[\sin\theta(t)+\frac{\ddot{\theta}(t)\cos\theta(t)}{g^2 + \dot{\theta}^2(t)}] & 0 \\
g\qty[\sin\theta(t)+\frac{\ddot{\theta}(t)\cos\theta(t)}{g^2 + \dot{\theta}^2(t)}] & 0 & g\qty[\cos\theta(t)-\frac{\ddot{\theta}(t)\sin\theta(t)}{g^2 + \dot{\theta}^2(t)}] \\
0 & g\qty[\cos\theta(t)-\frac{\ddot{\theta}(t)\sin\theta(t)}{g^2 + \dot{\theta}^2(t)}] & 0
\end{pmatrix}.
\end{align}

To ensure that the initial and final states of the SATD protocol match those of STIRAP, we impose boundary conditions:
\begin{align}
V(t)|_{t=0,T} = 0, \quad
H_\mathrm{ctrl}(t)|_{t=0,T} = 0, \quad
\theta(0) = 0, \quad
\theta(T) = \frac{\pi}{2}.
\end{align}
The lowest-order polynomial satisfying these boundary conditions is:
\begin{align}
\theta(t) = \frac{\pi}{2} \qty{6\qty(\frac{t}{T})^5 - 15\qty(\frac{t}{T})^4 + 10\qty(\frac{t}{T})^3}.
\end{align}

\section{Numerical simulation of the SATD-based state transfers on toy models of the $l$~coupler module}
\begin{figure*}
    \centering
	\includegraphics[width=\textwidth]{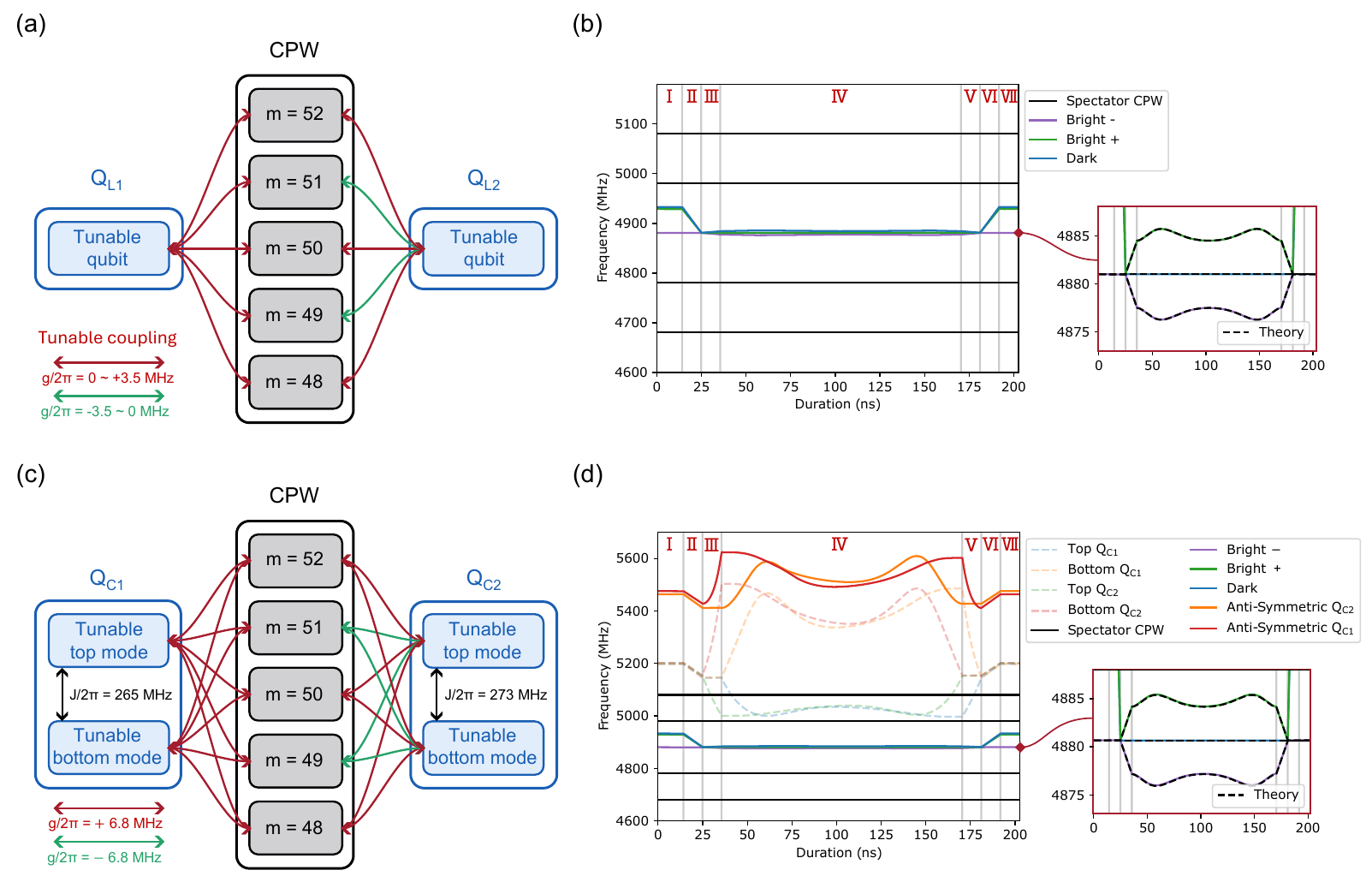}
    \caption{
    (a)
    Toy model 1: an $l$~coupler module consisting of tunable-frequency qubits with tunable couplings to a CPW, which is treated as a multi-mode resonator.
    (b)
    Eigenenergy diagram of the tunable-qubit $l$~coupler module during SATD-based state transfer.
    The horizontal and vertical axes represent time and eigenfrequencies, respectively.
    The horizontal purple line indicates the target CPW mode ($m=50$) used for the SATD-based state transfer, while the horizontal black lines represent the frequencies of the other CPW modes, referred to as spectator modes.
    Blue and green curves represent the tunable qubit frequencies of $Q_{\mathrm{L1}}$ and $Q_{\mathrm{L2}}$, respectively.
    Vertical black lines and red labels indicate distinct time regions, explained in the appendix.
    (c)
    Toy model 2: an $l$~coupler module consisting of tunable TCQs with fixed couplings to a CPW, treated as a multi-mode resonator.
    Each tunable TCQ consists of top and bottom qubit modes coupled with each other at a fixed rate.
    (d)
    Eigenenergy diagram of the TCQ-based $l$~coupler module during SATD-based state transfer.
    The horizontal and vertical axes represent time and eigenfrequencies, respectively.
    The horizontal purple line indicates the target CPW mode ($m=50$) used for the SATD-based state transfer.
    Red and orange curves correspond to the anti-symmetric modes of $Q_{\mathrm{C1}}$ and $Q_{\mathrm{C2}}$, respectively, while blue and green curves correspond to the symmetric modes ($l$~qubits) of $Q_{\mathrm{C1}}$ and $Q_{\mathrm{C2}}$.
    Light-colored curves represent the top and bottom transmon modes of the TCQs.
    }
    \label{fig:numerical_test}
\end{figure*}
In the following sections, we compare our experimental results with numerical simulations of state transfer on the TCQ-based $l$~coupler module.
Here, we describe our simulation procedures.
The simulated Hamiltonian consists of two TCQs and a single CPW interconnect.

To simulate the SATD-based state transfer, we prepared two types of toy models of the $l$~coupler module.
\Cref{fig:numerical_test}~(a) shows the simpler toy model named ``toy1'', which consists of two tunable-frequency $l$~qubits with tunable couplings to a CPW.
To save computational resources, we treat the CPW as a multi-mode resonator including CPW modes with the mode index $m\in[48,49,50,51,52]$ and restrict the Hamiltonian to the single-excitation manifold.
Therefore, the Hamiltonian of toy1 is written as:
\begin{align}
\frac{H_\mathrm{toy1}(t)}{\hbar}=\sum_{i=1,2} \omega_{qi}(t)\ket{q_i}\bra{q_i} + \sum_{m\in[48,52]} \omega_{cm}\ket{c_m}\bra{c_m} + \sum_{i=1,2}\sum_{m\in[48,52]}\qty(-1)^{m\delta_{i=2}} g_{ic}(t)\qty(\ket{q_i}\bra{c_m}+\ket{c_m}\bra{q_i})\;,
\label{eq:toy1}
\end{align}
where $\ket{q_i}$ and $\ket{c_m}$ represent the $i$th $l$~qubit and the $m$th~CPW mode with tunable frequencies $\omega_{qi}(t)$ and $\omega_{cm}$, respectively, which are coupled via the tunable coupling $g_{ic}(t)$.
Note that the sign of the voltage at the CPW ends of each CPW mode depends on the parity of the mode index $m$, which means that odd modes have opposite signs of coupling for the $l$~qubits.
The superscript $\delta_{i=2}$ corresponds to a Dirac delta function, being equal to 1 for the 2nd qubit ($i=2$) and zero otherwise.
\Cref{fig:numerical_test}~(b) shows the eigenenergy diagram of toy1 during the SATD-based state transfer.
The horizontal and vertical axes represent the duration and the eigenfrequencies, respectively.
The horizontal purple line represents the target CPW mode ($m=50$) frequency used for the SATD-based state transfer.
The horizontal black lines represent the CPW mode frequencies except for that of the target CPW mode.
We refer to these non-target CPW modes as spectator CPW modes.
Blue and green curves represent the frequencies of the tunable qubits $Q_{\mathrm{L1}}$ and $Q_{\mathrm{L2}}$, respectively.
Vertical black lines and red-colored labels indicate time regions and their labels.
Regions 1 and 7 correspond to buffers for preventing interference of adjacent control pulses.
In regions 2 and 6, we shift the $l$~qubit frequencies from their idle positions to the target CPW mode frequency while keeping the couplings between the $l$~qubits and the target CPW mode at $0$.
In regions 3 and 5, we linearly turn on or off the coupling between the receiver or emitter qubits, respectively.
Note that the $l$~qubits with variable coupling while in regions 3 and 5 should stay in the ground state under ideal conditions.
Region 4 is the main part of the state transfer, where we sweep the couplings between the $l$~qubits and the target CPW mode following the coupling formulas of STIRAP and SATD.
An enlarged plot in \cref{fig:numerical_test}~(b) shows the eigenenergies within the Hilbert subspace spanned by the $l$~qubits and the target CPW mode, where the black dotted line represents the theoretically expected three-body anti-crossing under SATD-based state transfer.
As shown in the enlarged plot, toy1 is in good agreement with the theoretical expectation.

\Cref{fig:numerical_test}~(c) shows a more realistic toy model named ``toy2'', which consists of two tunable-frequency TCQs with fixed couplings to a CPW.
Each TCQ consists of tunable-frequency top and bottom qubit modes coupled to each other with fixed coupling rates $J_\mathrm{TB1}=265~\mathrm{MHz}$ and $J_\mathrm{TB2}=273~\mathrm{MHz}$, respectively, which are inferred from our experiments.
Both top and bottom modes are coupled to the CPW modes with a coupling rate of $g_{c}=6.5~\mathrm{MHz}$, which is given from the design target value.
Note that the signs of the coupling rates are determined in the same way as in toy1.
Therefore, the Hamiltonian of toy2 is written as:
\begin{align}
\frac{H_\mathrm{toy2}(t)}{\hbar}=
&
\sum_{i=1,2}\qty[\omega_\mathrm{Ti}(t)\ket{T_i}\bra{T_i} +\omega_\mathrm{Bi}(t)\ket{B_i}\bra{B_i}+J_\mathrm{TBi}\qty(\ket{T_i}\bra{B_i}+\ket{B_i}\bra{T_i})]
+ \sum_{m\in[48,52]} \omega_\mathrm{cm}\ket{c_m}\bra{c_m}
\nonumber\\
&
 + \sum_{i=1,2}\sum_{m\in[48,52]}(-1)^{m\delta_{i=2}}g_{c}\qty(\ket{T_i}\bra{c_m}+\ket{c_m}\bra{T_i}+\ket{B_i}\bra{c_m}+\ket{c_m}\bra{B_i}) \;,
\end{align}
with the top $\ket{T_i}$ and bottom $\ket{B_i}$ mode frequencies $\omega_\mathrm{Ti}, \omega_\mathrm{Bi}$ of the $i$th TCQ.
Using \cref{eq:tunable_coupling_tcq}, we can determine the flux waveforms required for the state transfer operation to independently adjust the $l$~qubit (symmetric mode of TCQs) frequencies and the couplings between the $l$~qubits and the target CPW mode ($m=50$). Moreover, we can also determine the top and bottom mode frequencies of both TCQs from these formulas, which is what is set during the experiments.

\Cref{fig:numerical_test}~(d) shows the eigenenergy diagram of toy2 during the SATD-based state transfer.
The horizontal and vertical axes represent the duration and the eigenfrequencies, respectively.
The horizontal purple line represents the target CPW mode ($m=50$) used for the SATD-based state transfer.
Red and orange curves represent the anti-symmetric modes of $Q_{\mathrm{C1}}$ and $Q_{\mathrm{C2}}$, respectively.
Blue and green curves represent the symmetric modes ($l$~qubits) of $Q_{\mathrm{C1}}$ and $Q_{\mathrm{C2}}$, respectively.
Light-colored curves represent the top and bottom transmon modes of the TCQs.
As shown in the enlarged plot, toy2 is also in good agreement with the theoretical expectation.

\section{Comparison of the STIRAP and SATD}\label{sec:stirap_satd}
\begin{figure*}
    \centering
	\includegraphics[width=\textwidth]{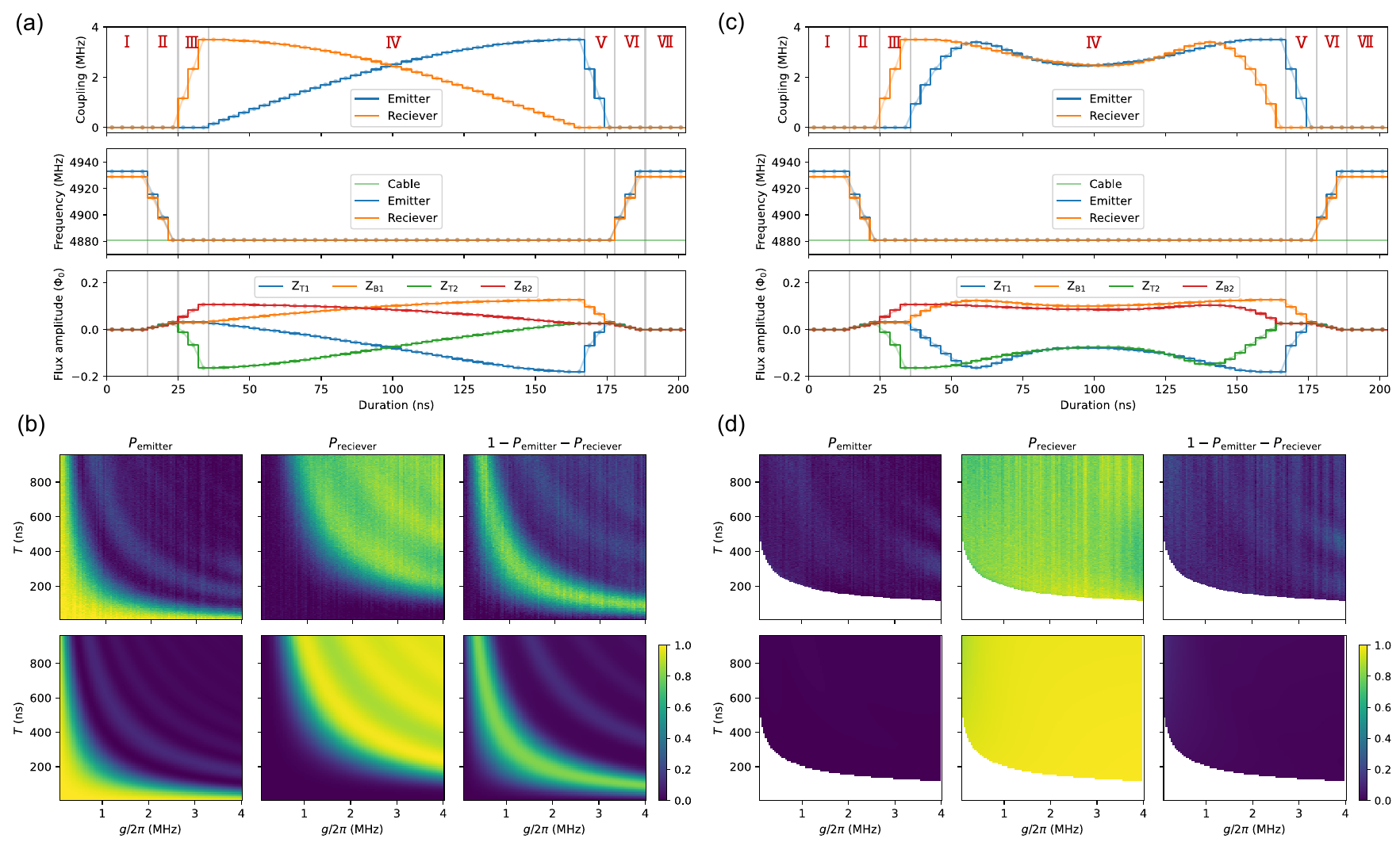}
    \caption{
    (a)
    The top and middle plots show the target coupling and frequency of the emitter $Q_\mathrm{L1}$ and receiver qubit $Q_\mathrm{L2}$ during the STIRAP-based state transfer with a total duration of $206~\mathrm{ns}$.
    The green horizontal line in the middle figure shows the target CPW mode frequency.
    Vertical black lines and red text represent time regions with different objectives and their labels, respectively.
    The bottom plot shows the flux pulse amplitudes used for the STIRAP-based state transfer.
    (b)
    Experimental results (upper) and Lindblad simulations (lower) of the STIRAP-based state transfer from the emitter $Q_\mathrm{L1}$ to the receiver qubit $Q_\mathrm{L2}$ while sweeping the parameters $g$ and $T$.
    Each figure in the upper or lower rows shows the population of the emitter $Q_{\mathrm{L1}}$, receiver $Q_{\mathrm{L2}}$, and the rest of the Hilbert space, respectively.
    (c)
    Same plots as in (a) but for the SATD-based state transfer.
    (d)
    Same plots as in (b) but for the SATD-based state transfer.
    }
    \label{fig:stirap_vs_satd}
\end{figure*}

SATD is an extension of the stimulated Raman adiabatic passage~(STIRAP)~\cite{Gaubatz_Population_1990, Vitanov_Stimulated_2017, Bergmann_Roadmap_2019, Chang_Remote_2020} with counter-diabatic corrections \cite{PhysRevLett.116.230503, zhou2017accelerated, PhysRevApplied.22.024006}, requiring modified control of the mixing angle.
For a single-mode system, SATD cancels out leakage to the bright states, hence speeding up the state transfer compared to STIRAP.
In a multimode system, the speed is further limited by leakage to adjacent interconnect modes, as shown in Ref.~\cite{PhysRevApplied.22.024006}.
The top and middle plots in \cref{fig:stirap_vs_satd}~(a) and (c) show the target time-domain Hamiltonian of the STIRAP- and SATD-based state transfers with a total duration of $206~\mathrm{ns}$ from emitter $Q_\mathrm{L1}$ to receiver $Q_\mathrm{L2}$, respectively.
The top plots show the target coupling strength between the emitter or receiver $l$~qubits and the target CPW mode.
The bottom plots show the target frequencies of the emitter and receiver $l$~qubits as blue and orange curves.
The green horizontal line represents the target CPW mode frequency.
The bottom plots in \cref{fig:stirap_vs_satd}~(a) and (c) show the actual flux pulse shapes used for the STIRAP- and SATD-based state transfer from $Q_{\mathrm{L1}}$ to $Q_{\mathrm{L2}}$, respectively.
The flux pulses are generated by arbitrary waveform generators with a time resolution of $3.56~\mathrm{ns}$ and are smoothed via digital and analog low-pass filters.
We note that the time-reversal pulse shapes implement state transfer in the reverse direction.

\Cref{fig:stirap_vs_satd}(b) and (d) show the population in the receiver qubit, emitter qubit, and the rest of the Hilbert space after the state transfer using STIRAP and SATD, respectively.
The upper and lower figures show the experimental and Lindblad simulation results, where we truncate the total excitation number of the system to $1$.
In each figure, we run a two-dimensional sweep of the parameters $g$ and $T$, corresponding to the maximum coupling and the execution time for STIRAP and SATD.
The experimental results are in qualitative agreement with the numerical simulations.
Under STIRAP, after the state transfer, the receiver qubit population significantly decreases in the region with faster execution $T$ and lower coupling $g$, which is caused by diabatic leakage from the dark mode to the bright modes during the state transfer.
On the other hand, under SATD, the receiver qubit population is not drastically suppressed in the same region.
Note that the counter-diabatic terms of SATD, the second terms of $g_\mathrm{r,e}$ in \cref{eq:coupling_satd}, become larger in the region with faster execution $T$ and lower coupling $g$.
In these demonstrations, however, we set the maximum coupling strength between the qubits and the target CPW mode to $4$~MHz.
The white-colored region in \cref{fig:stirap_vs_satd}~(d) represents the region where the required coupling strength exceeds $4$~MHz.

\section{Frame tracking}\label{sec:frame_tracking}
The quantum state of the $i$th qubit $Q_i$ is rotating at its eigenfrequency $\omega_i$ around the Pauli Z axis.
Since single-qubit gates are performed via microwave pulses resonant with the qubits, the initial phases of pulses on $Q_i$ at each time $t$ must follow the qubit phase $\phi_i^{(q)}(t)=\omega_i t$.
In addition to the qubit phase, the initial phases of the microwave pulses must also account for the history of the virtual Z rotations~\cite{PhysRevA.96.022330} applied to the qubit.
We define the sum of the qubit phase $\phi_i^{(q)}(t)$ and the accumulated virtual Z rotations $\phi_i^{(z)}(t)$ on $Q_i$ at time $t$ as the frame $\phi_i(t)$.
During the state transfer, we transfer not only the quantum state of the qubit but also the frame registered in the control software, which is part of the frame-tracking technique~\cite{PhysRevApplied.21.024018, PRXQuantum.5.020338}.

Let us consider the state transfer at a particular time $T$ from $Q_1$ to $Q_2$ with eigenfrequencies $\omega_{1,2}$, respectively.
After the state transfer, the initial frame of the emitter qubit $\phi^{\mathrm{before}}_1(T)$ is transferred to the receiver qubit $\phi^{\mathrm{after}}_2(T)$.
Because the emitter and receiver qubits have different eigenfrequencies, we reinterpret the accumulated phase of the virtual Z rotations on the receiver qubit as:
\begin{align}
\phi^{\mathrm{after}}_2(T)
&=\phi^{\mathrm{before}}_1(T) \nonumber\\
&=\phi_1^{(q)}(T) + \phi_1^{(z)}(T) \nonumber\\
&=\phi_2^{(q)}(T) + \phi_1^{(z)}(T) + \qty(\omega_1 - \omega_2)T\;.
\end{align}
Therefore, the accumulated virtual Z rotation on the receiver qubit $Q_2$ after the state tranfer is written as:
\begin{align}
\phi_2^{(z)}(T) = \phi_1^{(z)}(T) + \qty(\omega_1 - \omega_2)T\;.
\end{align}

\section{Error budget analysis of state transfer}

\begin{figure*}
    \centering
	\includegraphics[width=\textwidth]{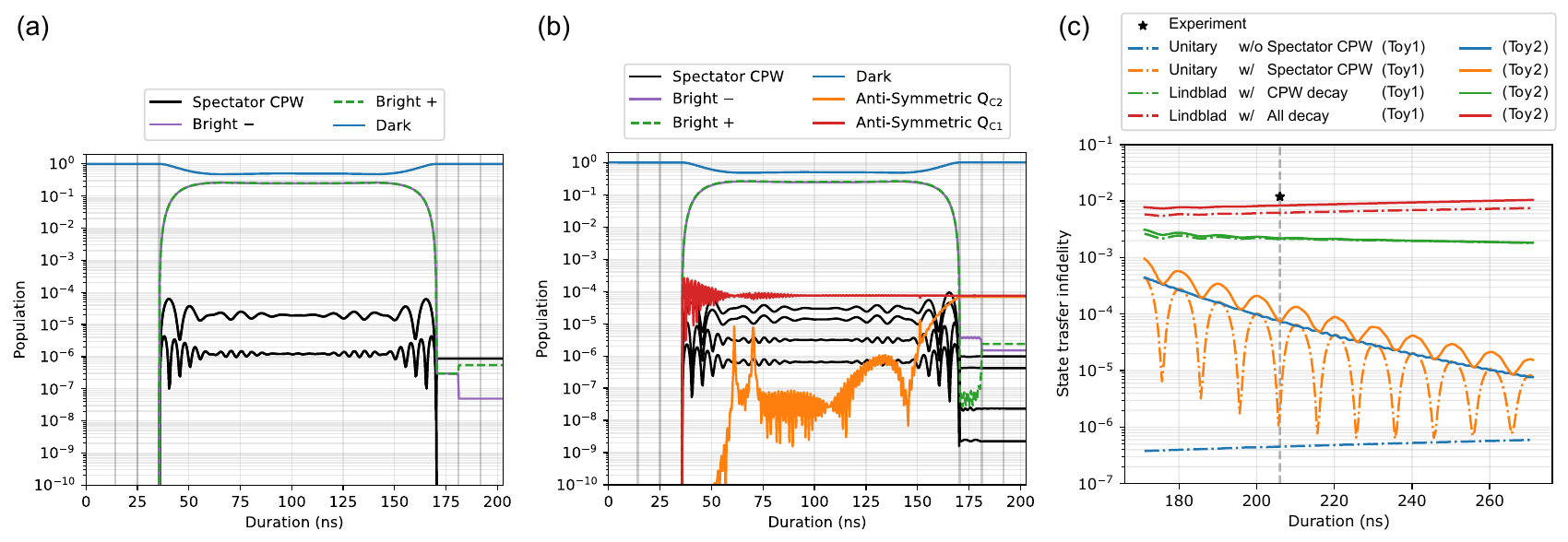}
    \caption{
    (a)
    Numerically calculated population of the instantaneous eigenstates during the SATD-based state transfer started from the excited state of the emitter qubit on toy model 1.
    (b)
    Numerically calculated population of the instantaneous eigenstates during the SATD-based state transfer started from the excited state of the emitter qubit on toy model 2.
    (c)
    Numerically calculated state transfer infidelities on the toy models while sweeping total duration of the SATD-based state transfer with various conditions.
    }
    \label{fig:error_budget}
\end{figure*}

In this section, we perform an analysis of the error budget of the SATD-based state transfer on the TCQ-based $l$~coupler module via numerical simulations.
\Cref{fig:error_budget}~(a) and (b) show the unitary numerical simulations of the SATD-based state transfer with the initial excited state in the emitter $l$~qubits for a total duration of $206~\mathrm{ns}$ on toy1 and toy2, respectively.
The horizontal and vertical axes represent the duration and the population of the instantaneous eigenstates during the state transfers.
The blue, green, and purple curves represent the population in the dark and bright states spanned by the $l$~qubits and the target CPW mode, respectively.
Black curves represent the population in the spectator CPW modes.
Red and orange curves in \cref{fig:error_budget}~(b) represent the anti-symmetric modes of TCQs $Q_{\mathrm{C1,2}}$, respectively.
As can be seen in the figures, the maximum error in state transfer in toy1 is leakage to spectator CPW modes of $1\times10^{-6}$, whereas in toy2 it is leakage into the antisymmetric modes of the TCQs of $1\times10^{-4}$, which occurs while sweeping the mixing angle between the top and bottom modes of the TCQs in the SATD-based state transfer.
Despite these errors, we emphasize that the simulation on toy2 shows a sufficiently low state transfer inefficiency of $2\times10^{-4}$.

We next calculate the state transfer infidelity on the toy models while sweeping the total duration of the state transfer in \cref{fig:error_budget}~(c).
The state transfer infidelity is calculated by averaging the fidelity between the initial state on the emitter $l$~qubit and the final state on the receiver $l$~qubit for the ground state, excited state, plus state, minus state, i-plus state, and i-minus state.
The horizontal and vertical axes represent the duration and the state transfer infidelity, respectively.
The dotted and solid curves represent the numerical calculation results on toy1 and toy2, respectively.
To clarify the error budget, we gradually added error sources to the numerical calculations.
Blue curves represent the unitary simulation where the CPW is approximated as a single-mode resonator consisting only of the target CPW mode ($m=50$).
In this simulation, leakage from the dark mode to the bright modes is the main source of error.
Since the toy1 Hamiltonian is equivalent to the ideal SATD Hamiltonian, leakage should be analytically zero.
However, numerical calculations show an error of approximately $1\times10^{-7}$ due to finite computational accuracy.
In the case of toy2, however, leakage actually occurs from the $l$~qubits to the anti-symmetric modes of the TCQs.
These antisymmetric modes also cause a time-dependent Lamb shift in the target CPW mode, resulting in a deviation from the ideal SATD.
State transfer infidelity of approximately $3\times10^{-4}$ appears under the experimental condition of duration $206~\mathrm{ns}$.
The orange curves represent the unitary simulation where the CPW is approximated as a multi-mode resonator ($m=[48,49,50,51,52]$), including spectator CPW modes.
Both toy1 and toy2 exhibit leakage to the spectator CPW modes, and the state transfer infidelity oscillates periodically with respect to the SATD-based state transfer duration.
From \cref{fig:error_budget}~(c), one can see that the leakage to the spectator CPW modes is negligibly smaller than that to the TCQ anti-symmetric modes at a duration of $206~\mathrm{ns}$.
The green curves represent Lindblad simulations including only the energy decay of the CPW modes, where the CPW is approximated as a multi-mode resonator ($m=[48,49,50,51,52]$).
The lifetimes of the CPW modes are taken from experimental values cited in \cref{Tab:device_parameters}.
Although the lifetimes of the CPW modes are much shorter than the lifetimes and the coherence times of the TCQs, the state transfer infidelity caused by CPW energy decay is only $2\times10^{-3}$ for both toy1 and toy2 at a duration of $206~\mathrm{ns}$.
This is because the SATD protocol suppresses excitation of the target CPW mode by using the dark state during the state transfer.
The larger error at shorter durations arises because the faster SATD protocol has a larger contribution of the target CPW mode in the dark state, as shown in \cref{eq:satd_frame_change}.
The red curves represent Lindblad simulations including the energy decay of the CPW modes and the energy decay and decoherence of the TCQs, where the CPW is approximated as a multi-mode resonator ($m=[48,49,50,51,52]$).
The state transfer infidelity at a duration of $206~\mathrm{ns}$ in toy1 and toy2 is $0.0062$ and $0.0083$, respectively.

The numerical calculations reveal that the main error source of the SATD-based state transfer on the TCQ-based $l$~coupler module is estimated to be energy decay and decoherence, which account for approximately $2/3$ of the experimental EPS.
The remaining approximately $3\%$ originates from leakage to the anti-symmetric modes of the TCQs.
Note that increasing the coupling strength between the TCQs and the CPW from $6.5$ to $15~\mathrm{MHz}$ should suppress both errors significantly while keeping the CPW length unchanged.

\section{Populations of the spectator qubits in the benchmarking experiments} \label{sec:spectator}
In \cref{fig:rb}~(c) and (d), we showed the population of the qubits, twirled by Clifford gates, in the benchmarking experiments.
However, there is a finite amount of population leakage to the spectator qubits.
In this appendix, we discuss the spectator qubit population during the benchmarking experiments.

\begin{figure*}
    \centering
	\includegraphics[width=\textwidth]{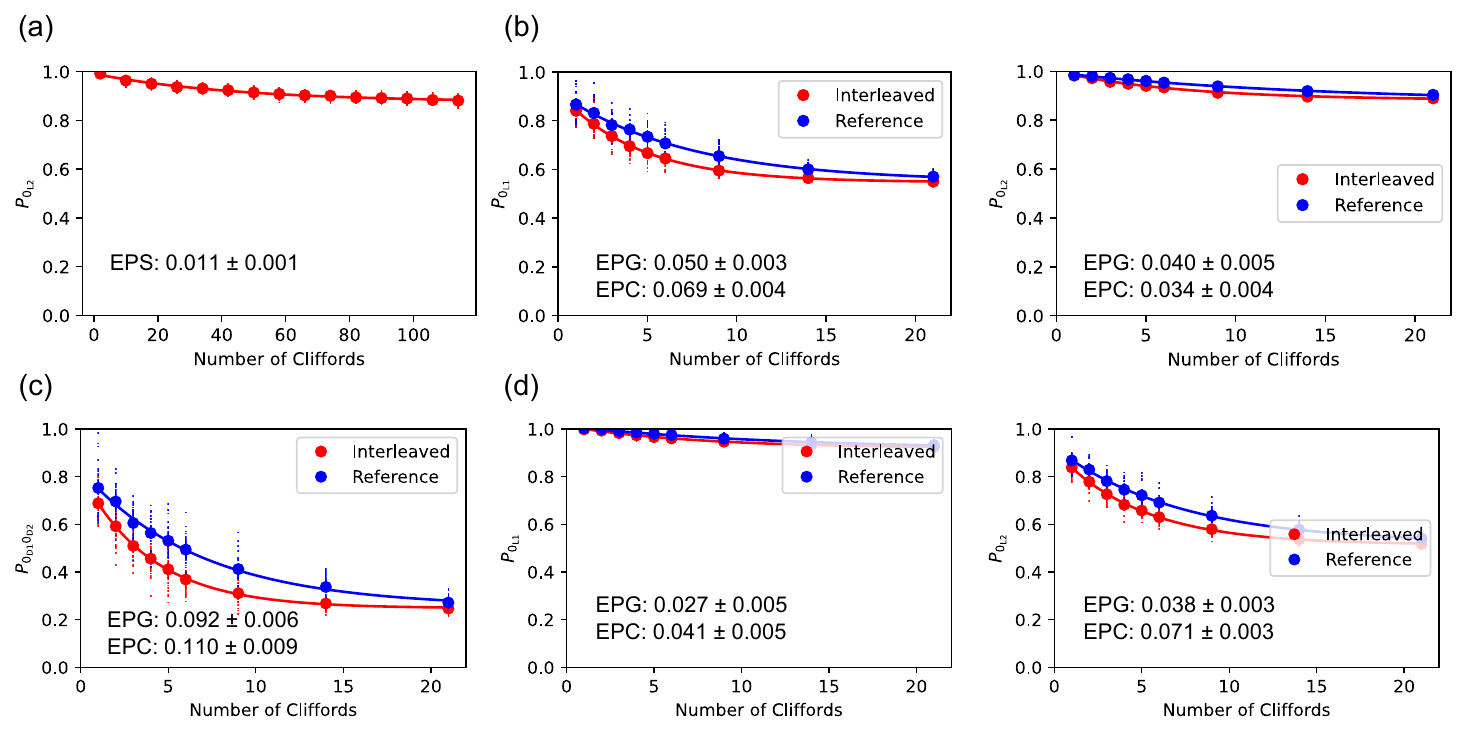}
    \caption{
    (a)
    Probability of measuring $Q_{\mathrm{L2}}$ in the $\ket{0}$ state as a function of the number of Cliffords in the network benchmarking.
    EPS represents the error per state transfer.
    (b)
    Probability of measuring $Q_{\mathrm{L1,2}}$ in the $\ket{0}$ state as a function of the number of Cliffords in the randomized benchmarking with the remote CNOT gate from $Q_\mathrm{D1}$ to $Q_\mathrm{D2}$.
    EPC and EPG represent the error per Clifford and the error per gate, respectively.
    (c)
    Two-qubit randomized benchmarking on the data qubits for the remote CNOT gate from $Q_\mathrm{D2}$ to $Q_\mathrm{D1}$: probability of measuring the data qubits $Q_\mathrm{D1,2}$ in the $\ket{00}$ state as a function of the number of Cliffords.
    We measure an EPG of $0.092$.
    (d)
    Probability of measuring $Q_{\mathrm{L1,2}}$ in the $\ket{0}$ state as a function of the number of Cliffords in the randomized benchmarking with the remote CNOT gate from $Q_\mathrm{D2}$ to $Q_\mathrm{D1}$.
    }
    \label{fig:spectator}
\end{figure*}

\Cref{fig:spectator}~(a) shows the probability of measuring $Q_\mathrm{L2}$ in the $\ket{0}$ state as a function of the number of Cliffords in the NB.
The probability decreases with the number of Cliffords and converges to $0.875$ rather than $0.5$.
As shown in \cref{eq:bright_dark}, at the beginning and the end of the state transfer protocol, the system eigenstates are ideally written as:
\begin{align}
\ket{D(0)}&=\ket{100}\;,\\
\ket{B_\pm(0)}&=\frac{1}{\sqrt{2}}\qty(\pm\ket{010}-\ket{001})\;,
\end{align}
and
\begin{align}
\ket{D(T)}&=-\ket{001}\;,\\
\ket{B_\pm(T)}&=\frac{1}{\sqrt{2}}\qty(\ket{100}\pm\ket{010})\;,
\end{align}
respectively.
Thus, the initial population of the receiver qubit $\ket{001}$ is transformed as:
\begin{align}
\ket{001}
&=-\frac{1}{\sqrt{2}}\qty(\ket{B_+(0)} + \ket{B_-(0)}) \nonumber\\
&\rightarrow -\frac{1}{\sqrt{2}}\qty(\ket{B_+(T)} + \ket{B_-(T)}) \nonumber\\
&=-\ket{100}\;,
\end{align}
which shows that the initial population in the emitter and receiver qubits is swapped.
By disentangling the population swap at each state transfer, it can be seen that the NB circuit is decomposed into a single qubit performing a single-qubit RB and a spectator qubit without running any gates.
The population increase in the spectator qubit is purely derived from leakage or decoherence in the state transfer.
Similar to leakage randomized benchmarking~\cite{PhysRevA.97.032306}, the random Clifford circuits in the NB twirl the leakage noise channel, which is why the population in the spectator qubit follows an exponential decay curve, where the decay rate represents the leakage rate $0.011\pm0.001$.

In \cref{fig:spectator}~(b), the left and right figures show the probability of measuring the spectator qubits $Q_\mathrm{L1,2}$ in the $\ket{0}$ state as a function of the number of Cliffords in the TQRB of the remote CNOT gate, respectively.
Both probabilities $p_{\mathrm{L1,2}}$ decrease with the number of Cliffords and converge to $0.550$ and $0.873$, respectively.
As shown in \cref{fig:rb}~(b), by disentangling the population swap at each state transfer, it can be seen that the remote CNOT gate uses only one of the $l$~qubits to propagate an excitation between the modules, and the other qubit always stays in the $\ket{0}$ state.
The main source of leakage error to the spectator qubits is thought to be decoherence acting on the intermediate excitation.
The difference in the asymptotic (large number of Cliffords) values of the probabilities between $Q_\mathrm{L1}$ and $Q_\mathrm{L2}$ is due to more intermediate excitation in $Q_\mathrm{L1}$ than in $Q_\mathrm{L2}$, and hence a higher leakage rate.
\Cref{fig:spectator}~(d) shows the same plots as \cref{fig:spectator}~(b), but for a remote CNOT gate from $Q_\mathrm{L2}$ to $Q_\mathrm{L1}$ shown in \cref{fig:spectator}~(c), where $Q_\mathrm{L2}$ experienced more errors than $Q_\mathrm{L1}$, which is contrary to what was shown in \cref{fig:spectator}~(b) and consistent with our analysis.

\section{Bell state preparation} \label{sec:bell}
\begin{figure*}
    \centering
	\includegraphics[width=\textwidth]{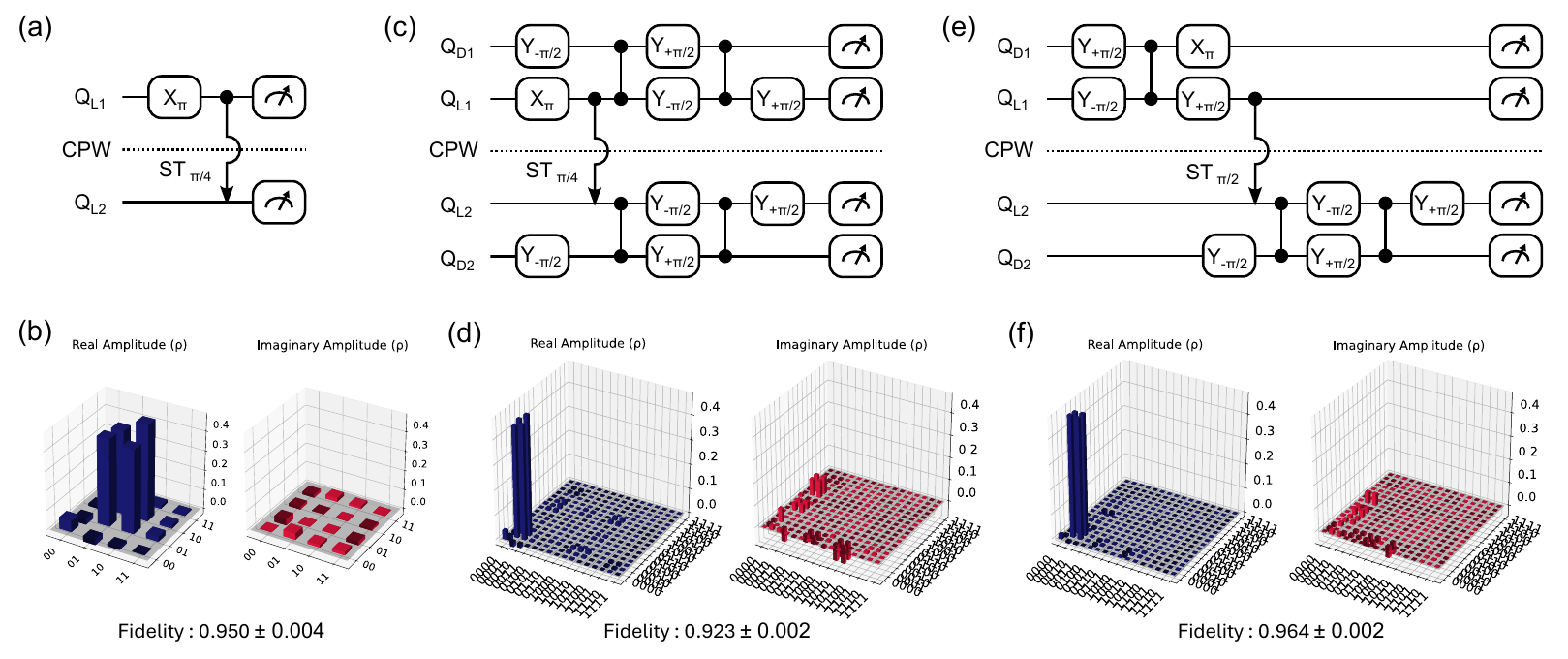}
    \caption{
    (a)
    Quantum circuit to prepare a Bell state on the $l$~qubits.
    (b)
    Density matrix of a prepared Bell state on the $l$~qubits using method (a).
    State labels follow the order of $Q_\mathrm{L1}$ and $Q_\mathrm{L2}$.
    (c)
    Quantum circuit to prepare a Bell state on the data qubits using a square-root of state transfer.
    (d)
    Density matrix of a prepared Bell state on the data qubits using method (c).
    State labels follow the order of $Q_\mathrm{D1}$, $Q_\mathrm{D2}$, $Q_\mathrm{L1}$, and $Q_\mathrm{L2}$.
    (e)
    Quantum circuit to prepare a Bell state on the data qubits using a full state transfer.
    (f)
    Density matrix of a prepared Bell state on the data qubits using method (e).
    State labels follow the order of $Q_\mathrm{D1}$, $Q_\mathrm{D2}$, $Q_\mathrm{L1}$, and $Q_\mathrm{L2}$.
    }
    \label{fig:bell}
\end{figure*}

In this section, we demonstrate Bell state preparation across the $l$~coupler module.
To prepare a Bell state distributed between data qubits, the most straightforward strategy is to use a remote CNOT gate between the data qubits.
However, there are more efficient ways to prepare Bell states on the data qubits in terms of circuit depth, as shown in \cref{fig:bell}~(b) and (c), where we use only a square-root of state transfer $ST_{\pi/4}$ or a full state transfer $ST_{\pi/2}$.
The square-root of state transfer is implemented via the modified SATD coupling sweep schedule as:
\begin{align}
\theta(t)&=\frac{\pi}{4}\qty{6\qty(\frac{t}{T})^5 - 15\qty(\frac{t}{T})^4 + 10\qty(\frac{t}{T})^3}\;.
\end{align}
Assuming equivalent execution times for the $ST_{\pi/4}$ and $ST_{\pi/2}$ operations, the $ST_{\pi/4}$ has slower coupling sweeps than $ST_{\pi/2}$ since it must regress to idle biases at the end of the coupling sweep schedule diabatically to suppress leakage of both $l$~qubit excitations while turning off the couplings.
Therefore, we compare these two methods below.

We first prepare a Bell state on the $l$~qubits using $ST_{\pi/4}$, where we adopt the same duration $T$ and maximum coupling $g$ as for the full state transfer in the main text.
\Cref{fig:bell}~(a) and (b) show a quantum circuit to prepare the Bell state and the density matrix of the prepared Bell state with a fidelity of $0.950\pm0.004$.
We next prepare a Bell state on the data qubits using $ST_{\pi/4}$ and local CZ gates.
\Cref{fig:bell}~(c) and (d) show the Bell state preparation circuit using $ST_{\pi/4}$ and the density matrix of the prepared Bell state with a fidelity of $0.923\pm0.002$.
Finally, we prepare a Bell state on the data qubits using $ST_{\pi/2}$ and local CZ gates.
\Cref{fig:bell}~(e) and (f) show the Bell state preparation circuit using $ST_{\pi/2}$ and the density matrix of the prepared Bell state with a fidelity of $0.964\pm0.002$.

\end{document}